\documentclass[12pt]{article}

\begin{document}

\title{ Hamiltonian analysis for  Lifshitz type Fields}

\author{ Alejandro Gaona $^{(1)}$\thanks{2122800633@alumnos.cua.uam.mx}, Juan M. Romero $^{(2)}$\thanks{jromero@correo.cua.uam.mx}\\
[0.3cm]\\
\it $^{(1)}$Posgrado en Ciencias Naturales e Ingenier\'ia,\\  
\it Universidad Aut\'onoma Metropolitana-Cuajimalpa,\\
\it M\'exico, D.F  05300, M\'exico\\
[0.5cm]
\it  $^{(2)}$Departamento de Matem\'aticas Aplicadas y Sistemas\\
\it Universidad Aut\'onoma Metropolitana-Cuajimalpa\\
\it M\'exico, D.F  05300, M\'exico\\
[0.5cm]}
\date{}

\pagestyle{plain}

\maketitle

\begin{abstract}
Using the Dirac  Method, we study the Hamiltonian consistency for three field theories. First we study the electrodynamics a  la Ho\v{r}ava and we show that this system is consistent for an arbitrary dynamical exponent $z.$ Second, we  study   a Lifshitz type electrodynamics, which was proposed in \cite{jean1:gnus}. For this last system we found that the canonical momentum and the electrical field are related through a Proca type Green function, however this system is consistent.   In addition, we show that  the anisotropic Yang-Mills theory with dynamical exponent  $z=2$ is consistent. Finally, we study a generalized  anisotropic Yang-Mills theory and we show that this last system is consistent too.

\end{abstract}

\section{Introduction}

Lately,  systems  invariant under anisotropic scaling
\begin{eqnarray}
\vec x \to b \vec x, \qquad t \to b^{z}t, \quad b={\rm constant} ,   \label{eq:1}
\end{eqnarray}
where $z$ is  a dynamical exponent, have been attracted a lot of  attention.  For example, in general relativity it has been found  space-times invariant under the anisotropic scaling (\ref{eq:1}), see \cite{marika:gnus}.  Some of  these kind of space-times can be seen as generalized Schr\"odinger space-time \cite{marika2:gnus} and other as an  $AdS$ deformation \cite{cheng:gnus,marika3:gnus}.   It is worth to mention   that different anisotropic space-times  are solution for the Einstein's equation with energy momentum tensor produced by a  Proca Field \cite{marika:gnus}, which is a  massive field. Now,  originally the anisotropic scaling (\ref{eq:1})  were found in condensed  matter \cite{marika4:gnus}. Amazingly,  the gravity/gauge 
correspondence   allows  a relation between different metrics invariant under the anisotropic scaling (\ref{eq:1}) and some condensed matter systems \cite{marika5:gnus,marika6:gnus,marika7:gnus}. Furthermore, almost every  field theory  can be  transformed  into a field theory invariant under the anisotropic scaling (\ref{eq:1}), this deformed field  theory improves its  high energy behavior 
\cite{anselmi:gnus,anselmi1:gnus,anselmi2:gnus,anselmi3:gnus,anselmi4:gnus,anselmi5:gnus,anselmi6:gnus}. In fact, using the anisotropic scalings (\ref{eq:1}),  Ho\v{r}ava formulated a modified gravity  which seems to be free ghosts and power counting renormalizable  for $z=3$ \cite{Horava:gnus}. However, applying   Dirac Method, it was shown that this gravity has dynamical 
inconsistencies \cite{Henneaux:gnus}, but  with the same  method were found healthy extensions \cite{Pujolas:gnus,pujolas2:gnus}. 
Ho\v{r}ava gravity has different interesting properties, some works about these properties  can be seen in 
\cite{horava1:gnus,horava2:gnus,horava3:gnus,horava4:gnus,horava5:gnus,horava6:gnus,horava7:gnus,horava8:gnus}. The Dirac Method is important to understand the anisotropic gravity, however there is not  a  study about Hamiltonian consistency  for anisotropic gauge  fields  as anisotropic  electrodynamics  or anisotropic Yang-Mills field. \\

In this paper, using the Dirac  Method, we show that the  anisotropic electrodynamics is consistent for an arbitrary dynamical exponent $z.$ In addition, we show that the usual Coulomb  gauge condition is correct for this system. Furthermore,  
we  study   the Hamiltonian formalism for a Lifshitz type electrodynamics. For this last system we found that the canonical momentum and the electrical field are related through a Proca type Green function, however  this system has two degrees of freedom and  is consistent. It is worth mentioning that this last system was proposed for  generating neutrino masses dynamically  \cite{jean1:gnus}. 
Moreover, using  the Dirac  Method again, we find that  the Hamiltonian formalism for the anisotropic Yang-Mills theory with   $z=2$ proposed in  \cite{anselmi6:gnus} is consistent.
Also, we study a generalized  anisotropic Yang-Mills theory and we show that this last system is consistent too.\\

This paper is organized as follow: in the section $2$ we study the formalism for the electrodynamics  a la Ho\v{r}ava; in the section $3$ we study  a Lifshitz type electrodynamics; in the section $4$ the anisotropic Yang-Mills  is studied and  in the section $5$  our summary is given. 

\section{ Anisotropic Electrodynamics and Hamiltonian analysis }

In this section we study the Hamiltonian formalism for the anisotropic electrodynamics, 
the action for this system  is given by \cite{anselmi2:gnus}
\begin{eqnarray}
S=   \int cdt d\vec x {\cal L} = \int dt d\vec x\left( E_{i}E_{i}-\frac{1}{2}F_{ij}f\left(\nabla^{2}\right) F_{ij}\right),
\label{eq:eh2}
\end{eqnarray}
where $f(x)=\sum_{z\geq 1}a_{z} x^{z-1}$ and 
\begin{eqnarray}
E_{i} = -\left (\partial _{t}A_{i} + \partial_{i}\phi \right), \quad B_{i} = \left(\vec \nabla \times \vec  A\right)_{i}, \quad F_{ij} = \partial_{i}A_{j}- \partial_{j}A_{i} = \epsilon_{ijk}B_{k} .
\end{eqnarray}
The case $z=2$ was first studied in \cite{fradkin:gnus}. \\

Now, from the action (\ref{eq:eh2}) we obtain 
\begin{eqnarray}
\pi^{i} &=& \frac{\partial \mathcal L}{\partial(\partial_{t }A_{i})}= -2\alpha E_{i},\\
\pi^{0} &=& \frac{\partial \mathcal L}{\partial(\partial_{t} \phi )}= 0.
\end{eqnarray}
Then we have the primary constraint 
\begin{eqnarray}
\chi_{1}=\pi^{0}(\vec x ,t) \approx 0.\label{eq:2}
\end{eqnarray}
In addition,  the  canonical Hamiltonian for this system is given by 
\begin{eqnarray}
{H}_{c} (t)&=& \int d\vec x \left(\pi^{i}\partial_{t}A_{i}-\mathcal{L} \right)\nonumber\\
&=&  \int d\vec x \left(\frac{1}{4} \pi_{i}\pi_{i} 
+ B_{i} f(\nabla^{2})B_{i} - \phi  \partial_{i}\pi^{i}\right).
\end{eqnarray}
According to Dirac Method \cite{dirac1:gnus}, all Hamiltonian constraint   does not evolve. Then, the constraint (\ref{eq:2})
must  satisfy 
\begin{eqnarray}
\dot \chi_{1} \approx 0,
\end{eqnarray}
namely
\begin{eqnarray}
\dot \chi_{1}&=&\left \{\pi^{0}( \vec x ,t),{H}_{c}(t)\right \}=\left \{\pi^{0}(\vec x ,t),\int d\vec y \left(\frac{1}{4} \pi_{i}\pi_{i} 
+  B_{i} f(\nabla^{2})B_{i} - A_{0} \partial_{i}\pi^{i}\right)\right\}\nonumber\\&=& \partial_{i}\pi^{i}(\vec x,t).
\end{eqnarray}
This equation  implies the new constraint  
\begin{eqnarray}
 \chi_{2}&=& \partial_{i}\pi^{i}(\vec x ,t) \approx 0.
\end{eqnarray}
According to Dirac Method \cite{dirac1:gnus},  this last  constraint
must satisfy 
\begin{eqnarray}
\dot \chi_{2} \approx 0,
\end{eqnarray}
namely
\begin{eqnarray}
\dot \chi_{2}&=&\left\{\partial_{i}\pi^{i}( \vec x ,t),{H}_{c}\right\}=\left\{\partial_{i}\pi^{i}(\vec x,t),\int d \vec y 
\left( B_{j} f(\nabla^{2})B_{j}\right)\right\}\nonumber\\
&=& \partial_{i} \epsilon_{ilj}\partial_{l}\left(f(\nabla^{2})B_{j}  \right) = 0.
\end{eqnarray}
Thus, there are  not   more constraints. In conclusion, the Hamiltonian constraints are given by
\begin{eqnarray}
\chi_{1}= \pi^{0}( \vec x,t) \approx 0,\qquad  \chi_{2}= \partial_{i}\pi^{i} ( \vec x ,t) \approx 0,\label{eq:4}
\end{eqnarray}
which satisfy  
\begin{eqnarray}
 \left\{\chi_{1}(\vec x ,t),\chi_{2}(\vec x^{\prime} ,t)\right\} = 0.
\end{eqnarray}
For this reason,   the constraints (\ref{eq:4}) are first class constraints and them  generate  gauge transformations for the system (\ref{eq:eh2}). Then,   the gauge freedom for this system is the same of the usual electrodynamics. In particular, this system has two degrees  of freedom and there are not dynamical inconsistencies. \\

Furthermore, the extended Hamiltonian is given by
\begin{eqnarray}
{H}_{E}(t) &=&   \int d \vec x \left(\frac{1}{4} \pi_{i}\pi_{i} 
+  B_{i} f(\nabla^{2})B_{i} - A_{0} \partial_{i}\pi^{i}+\lambda_{1}\pi^{0}
+\lambda_{2}\partial_{i}\pi^{i}\right), \nonumber
\end{eqnarray}
 where $\lambda_{1}$ and  $\lambda_{2}$ are Lagrange multipliers.
Using this Hamiltonian we have the equations of motion 
\begin{eqnarray}
\dot \phi (\vec x,t)&=& \left\{ \phi (\vec x,t),H_{E}(t)\right\}= \lambda_{1}(\vec x ,t),\\
\dot A_{i}(\vec x ,t)&=& \left\{A_{i}( \vec x,t),H_{E}(t)\right\}=\frac{1}{2\alpha} \pi_{i}(\vec  x,t)
+ \partial_{i}\left[ \phi (\vec x,t)-\lambda_{2}(\vec x,t) \right], \qquad \\
\dot \pi ^{0}&=&\left\{\pi^{0}(\vec x,t),H_{E}(t)\right\}=\partial_{i}\pi^{i}( \vec x ,t)= 0, \\
\dot \pi_{i}&=&\left \{\pi_{i}(\vec x,t),H_{E}(t)\right\} =\epsilon_{ilj}\partial_{l}\left(f(\nabla^{2})B_{j}(\vec x,t)\right).
\end{eqnarray}
The two last equations can be written as  
\begin{eqnarray}
\nabla \cdot \vec E&=&0,\\
{\vec{\nabla}}\times \big(f(\nabla^{2}) {\vec{B}} \big) &=& \frac{\partial \vec{E}}{\partial t},
\end{eqnarray}
which are the equations  of motion for anisotropic electrodynamics \cite{anselmi2:gnus}. \\

\section{ Coulomb gauge}

Due  that this theory has two first class constraints, we have to choice two gauge conditions. 
The  usual  Coulomb gauge conditions  are 
\begin{eqnarray}
\chi_{3}(\vec x,t) &=& \phi (\vec x,t) \approx 0, \label{eq:5}\\
\chi_{4}(\vec x,t) &=& \partial_{i} A^{i}(\vec x ,t) \approx 0 \label{eq:6},
\end{eqnarray}
which are good gauge conditions for the anisotropic electrodynamics. In fact, according to  Dirac  Method, the  constraints 
 (\ref{eq:5})-(\ref{eq:6}) have not evolve. Then, 
\begin{eqnarray}
\dot \phi (\vec x,t)&=& \left\{ \phi (\vec x,t),H_{E}(t)\right\}= \lambda_{1}(\vec x ,t) \approx 0,\\
\dot \chi_{4}&=&\left\{\partial_{i}A_{i},H_{E}\right\} =\left\{ \partial_{i}A_{i},H_{c} + \int d\vec y \left(\lambda_{1}\pi^{0} 
+\lambda_{2}\partial_{j}\pi^{j}\right) \right\}\nonumber\\
&=&  \frac{1}{2\alpha} \partial_{i}\pi_{i}(\vec x )
+ \nabla^{2}\phi 
- \partial_{i}\partial_{i}\lambda_{2} \approx 0 \nonumber,
\end{eqnarray}
which implies 
\begin{eqnarray}
\nabla^{2}\phi - \partial_{i}\partial_{i}\lambda_{2} \approx 0.
\end{eqnarray}
Now, the  Dirac  bracket   for this gauge is 
\begin{eqnarray}
& & \left\{V(\vec x,t),W(\vec y,t)\right\}^{*}=\left\{V(\vec x,t),W(\vec y,t)\right\} \nonumber \\
&&+\int d \vec y^{\prime} \left[ \left\{V,\pi^{0}(\vec y^{\prime},t) \right\} \left\{A^{0}(\vec y^{\prime},t),W \right\} - \left\{V, A^{0}(\vec y^{\prime} ,t)\right\} \left\{\pi^{0}(\vec y^{\prime} ,t),W\right\} \right] \nonumber \\ 
&&-\int \int \frac{1}{4\pi} \frac{1}{\mid \vec w -\vec y^{\prime } \mid} \left \{V,\partial_{i}\pi^{i}(\vec w,t)\right\} 
\left\{\partial_{i}A^{j}(\vec y^{\prime}, t),W\right\}   d \vec y^{\prime} d \vec w \nonumber \\
&&+\int \int \frac{1}{4\pi}\frac{1}{\mid \vec w- \vec y^{\prime} \mid} \left\{V, \partial_{j}A^{j}(\vec w, t)\right\} 
\left\{\partial_{i}\pi^{i}( \vec y^{\prime} , t),W\right \} d\vec y^{\prime}  d \vec w. \nonumber 
\end{eqnarray}
In particular,  we have 
\begin{eqnarray}
\left\{A^{\mu} (\vec x,t),\pi^{\nu}(\vec y,t ) \right\}^{*}&=&\left(\eta^{\mu\nu} + \eta^{\mu 0}\eta^{0\nu} \right)\delta^{3}(\vec x-\vec y) \nonumber\\
& -&\frac{\partial}{\partial x_{\mu}} \frac{\partial}{\partial y_{\nu}} \left( \frac{1}{4\pi} \frac{1}{\mid \vec x-\vec y \mid} \right). 
\end{eqnarray}
Then,  the equations of motion are 
\begin{eqnarray}
\left\{A^{i} (\vec x,t ),H_{E}(t)\right\}^{*} = \dot{A}^{i},\\
\left\{\pi^{i} (\vec x,t ),H_{E}(t) \right\}^{*} =  \vec{\nabla}\times \left(f(\nabla^{2})\vec{B}\right) =  \frac{\partial \vec{E}}{\partial t}.
\end{eqnarray}
Thus,  the usual Coulomb gauge is a good gauge condition for the system (\ref{eq:eh2}).

\section{Lifshitz type electrodynamics }

Recently it was proposed a new mechanism  to obtain massive fields. In this mechanism  spatial higher order derivatives 
are introduced \cite{jean1:gnus}. For example, the action for the Lifshitz type electrodynamics is 
\begin{eqnarray}
S&=& \int d\vec x dt {\cal  L}=   \int d\vec x dt \left( -\frac{1}{4} F_{\mu \nu}\left( 1- \frac{ \nabla^{2} }{M^{2}}  \right) F^{\mu \nu} \right), \\
&=&  \int d\vec x dt \left( \frac{1}{2} E_{i}E^{i} - E_{i} \frac{\nabla^{2} }{2M^{2} } E^{i} - \frac{1}{4} F_{ij}  F_{ij} + 
  F_{ij}\frac{\nabla^{2}}{4M^{2}}   F_{ij}  \right).
\end{eqnarray}
From this action we have 
\begin{eqnarray}
\pi^{0}&=&0, \\
\pi^{i }&=& -\left(1- \frac{\nabla^{2} }{M^{2}}  \right) E^{i}. \label{eq:7}
\end{eqnarray}
Then,  we have the primary constraint
\begin{eqnarray}
\chi_{1}=\pi^{0}(\vec x ,t) \approx 0.\label{eq:10}
\end{eqnarray}
In addition,  from the equation (\ref{eq:7}), we have
\begin{eqnarray}
E^{i}(\vec x, t)= \int d\vec x^{\prime} G\left(\vec x- \vec x^{\prime} \right) \pi^{i} (\vec x^{\prime}, t), 
\end{eqnarray}
where 
\begin{eqnarray}
 -\left(1- \frac{\nabla^{2} }{M^{2}}  \right)   G\left(\vec x- \vec x^{\prime} \right) = \delta^{3} (\vec x-\vec x^{\prime}),
\end{eqnarray}
namely
\begin{eqnarray}
 G\left(\vec x- \vec x^{\prime} \right) = -\frac{M^{2}}{4\pi } \frac{ e^{-M |\vec x- \vec x^{\prime}|}} { |\vec x- \vec x^{\prime} |}.
 \label{eq:procaa}
\end{eqnarray}
Notice that this Green function is a Proca type propagator. \\

Now,  the  canonical Hamiltonian for this system is given by 
\begin{eqnarray}
{H}_{c} (t)&=& \int d\vec x(\pi^{i}\partial_{t}A_{i}-\mathcal{L})\nonumber\\
&=&  \int d\vec x \left(-\frac{1}{2} \pi^{i}E_{i} +\frac{1}{4} F_{ij}F^{ij}-\frac{1}{4M^{2} } F^{ij} \Delta F_{ij}+ \phi  \partial_{i} \pi^{i}\right) \nonumber \\
&=&  \int d\vec x  d\vec x^{\prime} \left(-\frac{1}{2} \pi^{i}(\vec x,t)   G\left(\vec x- \vec x^{\prime} \right) \pi_{i} (\vec x^{\prime}, t) \right)\nonumber \\
 & &+\int d\vec x  \left( \frac{1}{4} F_{ij}F^{ij}+\frac{1}{4M^{2} } F^{ij} \Delta F_{ij}+\phi \partial_{i} \pi^{i}\right).
\end{eqnarray}
In addition, according  to the Dirac  Method \cite{dirac1:gnus}, the constraint (\ref{eq:10})
has  to satisfy 
\begin{eqnarray}
\dot \chi_{1} \approx 0,
\end{eqnarray}
namely
\begin{eqnarray}
\dot \chi_{1}&=&\left\{\pi^{0}( \vec x ,t),{H}_{c}(t)\right\}= \partial_{i}\pi^{i}(\vec x,t)\approx 0.
\end{eqnarray}
This equation  implies the new constraint  
\begin{eqnarray}
 \chi_{2}&=& \partial_{i}\pi^{i}(\vec x ,t) \approx 0, \label{eq:eh11}
\end{eqnarray}
which implies
\begin{eqnarray}
\dot \chi_{2} \approx 0.
\end{eqnarray}
It can be shown that 
\begin{eqnarray}
\dot \chi_{2}&=&\left\{\partial_{i}\pi^{i}( \vec x ,t),{H}_{c}(t)\right\} = 0.
\end{eqnarray}
Thus, there are  not   more constraints. Then,  all  the Hamiltonian constraints are given by the constraints (\ref{eq:10}) and 
(\ref{eq:eh11}) which  are first class constraints and   generate the gauge transformations for this system. For that reason   there are not dynamical inconsistencies.\\

Now,  the usual Proca field is massive  and has three degrees  of freedom. In   the  Lifshitz type electrodynamics appears  a  Proca type propagator (\ref{eq:procaa}),
however   it has  two first class constraints. For this last reason,  the  Lifshitz type electrodynamics   has two degrees  of freedom, as the usual electrodynamics.  \\

\section{Anisotropic Yang-Mills field }

The action for the anisotropic Yang-Mills field is given by \cite{anselmi6:gnus}
\begin{eqnarray}
\label{ymaniso1}
S &=& \frac{1}{4}\int dt d\vec x  \left( \frac{1}{e^{2}} (E_{ai}E_{ai})
+\beta(D_{i}F_{aik} D_{j} F_{ajk}) \right)  , \label{eq:h1}
\end{eqnarray}
where 
\begin{eqnarray}
D_{i}F_{ajk}= \partial_{i}F_{ajk}+igf_{ab}\-^{c}A_{bi}F_{cjk}. 
\end{eqnarray}
This action is invariant under anisotropic scaling  for $z=2.$\\

From the action (\ref{eq:h1}) we have
\begin{eqnarray}
\pi_{a}^{i} &=&-\frac{1}{2e^{2}}E^{i}_{a},\\
\pi_{a}^{0}& =& 0.
\end{eqnarray}
Then,   we have the constraints 
\begin{eqnarray}
\chi_{1a}=\pi_{a}^{0}& \approx& 0. \label{eq:ym1}
\end{eqnarray}
In addition, the canonical Hamiltonian is 
\begin{eqnarray}
{H}_{c} (t)=  \int d\vec x \left(e^{2}\pi_{a}^{i}\pi_{a}^{i} -\frac{\beta}{4} \left(D_{i}F_{aik} D_{j} F_{ajk}\right) 
- A_{a0}\left(\partial_{i}\pi_{a}^{i} +igf_{ac}\-^{b}A_{ci}\pi_{b}^{i} \right)\right).
\end{eqnarray}
Using this Hamiltonian we get 
\begin{eqnarray}
\dot \chi_{1a}=\{\pi_{a}^{0}(\vec x,t),{H}_{c}(t)\}=\partial_{i}\pi_{a}^{i}(\vec x,t).
+igf_{ae}\-^{b}\pi_{b}^{i}(\vec x,t)A_{ei}(\vec x,t)
\end{eqnarray}
Now, due that the Dirac Method sets 
\begin{eqnarray}
\dot \chi_{1a}\approx 0,
\end{eqnarray}
we have the new constraints
\begin{eqnarray}
\chi_{2a}(\vec x,t)=\partial_{i}\pi_{a}^{i}(\vec x,t)
+igf_{ae}\-^{b}\pi_{b}^{i}(\vec x,t)A_{ei}(\vec x,t)\approx 0, \label{eq:1y}
\end{eqnarray}
which satisfy
\begin{eqnarray}
\left\{\chi_{2a}(\vec x,t),\chi_{2b}(\vec y,t)\right\}= ig f_{ab}\-^{c} \chi_{2c}(\vec y,t) \approx 0.
\label{eq:v31}
\end{eqnarray}
Using the constraints (\ref{eq:1y}) we obtain 
\begin{eqnarray}
{H}_{c} (t)&=& \
 \int d\vec x \left(e^{2}\pi_{a}^{i}\pi_{a}^{i} -\frac{\beta}{4} \left(D_{i}F_{aik} D_{j} F_{ajk}\right) 
- A_{a0}\chi_{2a}\right).
\end{eqnarray}
Furthermore we found that
\begin{eqnarray}
\dot \chi_{2a}&=&\left\{\chi_{2a},{H}_{c}\right\}=-\frac{ig\beta}{2} f_{ad}\-^{h}\left(D_{i}F_{h}^{ik})(D_{j}F_{d}^{jk}\right)=0.
\end{eqnarray}
Then, there are not more constraints and  the only constraints for this system  are (\ref{eq:ym1}) and (\ref{eq:1y}), which are first class constraints.  This last result implies that the extended Hamiltonian is given by
\begin{eqnarray}
{H}_{E} (t)&=&  \int d\vec x \Bigg(e^{2}\pi_{a}^{i}\pi_{a}^{i} -\frac{\beta}{4} \left(D_{i}F_{aik} D_{j} F_{ajk}\right) 
+\lambda_{1a} \chi_{1a} + \left( \lambda_{2a}- A_{a0} \right)\chi_{2a}  \Bigg), \qquad 
\end{eqnarray}
where  $\lambda_{1a}$ and $\lambda_{2a}$ are Lagrange multipliers.\\

It is worth mentioning that  the constraints (\ref{eq:ym1}) and (\ref{eq:1y}) are the same constraints for the usual Yang-Mills theory \cite{dirac2:gnus}. Now, the consistent Coulomb gauge conditions for the usual Yang-Mills theory  are  given by \cite{dirac2:gnus}
\begin{eqnarray}
\partial_{i}A_{a}^{i}(\vec x,t) &=& 0,\nonumber\\
A_{a}^{0}(\vec x,t)-\frac{1}{4\pi}\int d\vec y {\bf G}_{ab}\left(\vec x,\vec y,A \right)(2e^{2})
igf_{bc}\-^{d}\pi_{d}^{i}(\vec y,t)A_{ci}(\vec y,t)   &\approx& 0, \nonumber
\end{eqnarray}
where 
\begin{eqnarray} 
\Big(\delta_{ab}\partial_{i}\partial^{i} 
+igf_{ac}\-^{b} A_{c}^{i}\partial_{i}\Big){\bf G}_{cb}(\vec x,\vec y,A) =-4\pi \delta_{ab} \delta(\vec x-\vec y).
\end{eqnarray}
It is possible to show that these gauge conditions are good gauge conditions for the anisotropic Yang-Mills with $z=2.$\\

\subsection{General case}

For the case $z=2,$ an alternative action for the anisotropic Yang-Mills theory is given by 
\begin{eqnarray}
S &=& \frac{1}{4}\int dt d\vec x  \left( \frac{1}{e^{2}} E_{ai}E_{ai}
+\beta F_{ajk} D^{2} F_{ajk}  \right) , \quad D^{2}F_{ajk}=D_{i}D_{i}F_{ajk}.
\end{eqnarray}
In fact, we can propose the general action 
\begin{eqnarray}
S &=& \frac{1}{4}\int dt d\vec x  \left( \frac{1}{e^{2}} E_{ai}E_{ai}
+F_{ajk} f\left(D^{2} \right) F_{ajk}) \right), \label{eq:h2a}
\end{eqnarray}
where $f(x)=\sum_{z\geq 1}a_{z} x^{z-1}.$ From this  action we have
\begin{eqnarray}
\pi_{a}^{i} &=&-\frac{1}{2e^{2}}E^{i}_{a},\\
\pi_{a}^{0}& =& 0.
\end{eqnarray}
Namely,   we obtain  the constraints 
\begin{eqnarray}
\chi_{1a}=\pi_{a}^{0}& \approx& 0. \label{eq:ym2}
\end{eqnarray}
In addition, the canonical Hamiltonian is 
\begin{eqnarray}
{H}_{c} (t)=  \int d\vec x \left(e^{2}\pi_{a}^{i}\pi_{a}^{i} -\frac{1}{4} F_{ajk} f\left(D^{2}\right)  F_{ajk} 
- A_{a0} \left(\partial_{i}\pi_{a}^{i} +igf_{ac}\-^{b}A_{ci}\pi_{b}^{i} \right)\right).
\end{eqnarray}
Now, using this Hamiltonian we arrive to 
\begin{eqnarray}
\dot \chi_{1a}=\left\{\pi_{a}^{0}(\vec x,t),{H}_{c}(t) \right\}=\partial_{i}\pi_{a}^{i}(\vec x,t)
+igf_{ae}\-^{b}\pi_{b}^{i}(\vec x,t)A_{ei}(\vec x,t), 
\end{eqnarray}
which implies the constraints 
\begin{eqnarray}
\chi_{2a}(\vec x,t)=\partial_{i}\pi_{a}^{i}(\vec x,t)
+igf_{ae}\-^{b}\pi_{b}^{i}(\vec x,t)A_{ei}(\vec x,t)\approx 0.
\end{eqnarray}
Due that the constraints  $\chi_{1a}(\vec x,t)$ and $\chi_{2a}(\vec x,t)$ satisfy (\ref{eq:v31}),  these  are first class constraints.\\ 

Now, using the Jacobi's identity for the structure constants
\begin{eqnarray}
 f_{rs}\-^{c}f_{ac}\-^{b}+f_{sa}\-^{c}f_{rc}\-^{b}+f_{ar}\-^{c}f_{sc}\-^{b}=0,\nonumber
\end{eqnarray}
we arrive to
\begin{eqnarray}
& &\left \{ D_{i}\pi_{a}^{i}(\vec x, t),  \int d\vec y F_{bjk}(\vec y,t) \left( D^{2}\right )^{z} F_{bjk}(\vec y,t)\right \} \nonumber\\
&=& 2(ig)^{2}  A_{r}^{j}(\vec x,t) A_{s}^{k} (\vec x,t) \left( f_{rs}\-^{c}f_{ac}\-^{b}+f_{sa}\-^{c}f_{rc}\-^{b}+f_{ar}\-^{c}f_{sc}\-^{b} \right)\left( D^{2}\right )^{z} F_{bjk}(\vec x,t) =0.
\qquad 
\end{eqnarray}
This last result implies 
\begin{eqnarray}
\dot \chi_{2a}&=&\{\chi_{2a},{H}_{c}\} \approx  0.
\end{eqnarray}
Then, there are not more constraints and  this system is consistent. 
In this case the extended Hamiltonian is given by
\begin{eqnarray}
{H}_{E} (t)&=&  \int d\vec x \Bigg(e^{2}\pi_{a}^{i}\pi_{a}^{i} -\frac{1}{4}  
F_{ajk} f \left( D^{2}\right) F_{ajk}
+\lambda_{1a} \chi_{1a} + \left( \lambda_{2a}- A_{a0} \right)\chi_{2a}  \Bigg), \qquad 
\end{eqnarray}
where  $\lambda_{1a}$ and $\lambda_{2a}$ are Lagrange multipliers.  Another generalized  anisotropic Yang-Mills theory was proposed in \cite{anselmi7:gnus}. \\

The Hamiltonian constraints for anisotropic gravity are very different   from the Hamiltonian constraints for the usual gravity.
However, the Hamiltonian constraints obtained for the Lifshitz type fields are not different from the Hamiltonian constraints for usual fields.
This result is interesting, because  if  it is possible to obtain a quantum  gravity with anisotropic scaling transformations, 
the anisotropic fields theories will not have problems, in fact them improve its UV behavior. 
  
\section{Summary} 
In  this paper, we studied the  dynamical consistency for the electrodynamics a la Ho\v{r}ava and we show that this system is consistent for arbitrary dynamical exponent $z.$ In fact, for this system the constraints are the same that the usual electrodynamics. For this reason, a good gauge condition for the usual electrodynamics is a good gauge condition for the anisotropic electrodynamics. In addition, we  study a Lifshitz type electrodynamics, which was proposed in \cite{jean1:gnus}. For this last system we found that the canonical momenta and electrical field are related through a Proca type Green function.  Also, we show that this last  system is consistent. 
The anisotropic Yang-Mills field was studied too, in this case  we show that  the anisotropic Yang-Mills theory with dynamical exponent  $z=2$ proposed in  
\cite{anselmi6:gnus}  is dynamical consistent.  Finally, we  studied  a generalized  anisotropic Yang-Mills theory and it was  shown     that this system  is consistent too.

\end{document}